# GPU accelerated Nature Inspired Methods for Modelling Large Scale Bi-Directional Pedestrian Movement

Sankha Baran Dutta, Robert McLeod, and Marcia Friesen
*Electrical and Computer Engineering*
*University of Manitoba*
*Winnipeg, Canada*
Email: umdutta@cc.umanitoba.ca, Robert.McLeod@ad.umanitoba.ca, Marcia.Friesen@ad.umanitoba.ca

*Abstract*— Pedestrian movement, although ubiquitous and well-studied, is still not that well understood due to the complicating nature of the embedded social dynamics. Interest among researchers in simulating pedestrian movement and interactions has grown significantly in part due to increased computational and visualization capabilities afforded by high power computing. Different approaches have been adopted to simulate pedestrian movement under various circumstances and interactions. In the present work, bi-directional crowd movement is simulated where an equal numbers of individuals try to reach the opposite sides of an environment. Two movement methods are considered. First a Least Effort Model (LEM) is investigated where agents try to take an optimal path with as minimal changes from their intended path as possible. Following this, a modified form of Ant Colony Optimization (ACO) is proposed, where individuals are guided by a goal of reaching the other side in a least effort mode as well as a pheromone trail left by predecessors. The basic idea is to increase agent interaction, thereby more closely reflecting a real world scenario. The methodology utilizes Graphics Processing Units (GPUs) for general purpose computing using the CUDA platform. Because of the inherent parallel properties associated with pedestrian movement such as proximate interactions of individuals on a 2D grid, GPUs are well suited. The main feature of the implementation undertaken here is that the parallelism is data driven. The data driven implementation leads to a speedup up to 18x compared to its sequential counterpart running on a single threaded CPU. The numbers of pedestrians considered in the model ranged from 2K to 100K representing numbers typical of mass gathering events. A detailed discussion addresses implementation challenges faced and averted. Detailed analysis is also provided on the throughput of pedestrians across the environment.

## I. Introduction

The study of pedestrian dynamics has attracted considerable research and interest has increased significantly over time. Pedestrian flow related issues arise in different instances such as in mass-gatherings, sporting events and even cross-walks within large urban centers. As the density of the crowd increases, the vulnerability towards an adverse event or disaster also increases. Replicating the situation in the real world to create a safer environment is not practical. As such, simulation has become an attractive alternative for simulating crisis situations when crowds are very large. However, simulating life-like crowd dynamics is an arduous task. Several models have been proposed to emulate crowd movements including social force models [1],[2],[3], cellular automata models [4],[5],[6], and gas-kinetic model [7], each of these have met with various degrees of success.

The work here emulates a situation where two groups of pedestrians are placed at opposite sides of an environment. Their goal is to reach the other side of the environment within a given number of time steps. To model this situation, two algorithms are used: The Least Effort Model (LEM) [18] and the Ant Colony Optimization (ACO) [9],[10]. The objective of the work is to compare the results of LEM and ACO simulations for multiple scenarios, both run on a GPU platform. Furthermore, an objective of the work is to compare the results of ACO simulations for multiple scenarios, run on both GPU and CPU platforms, as an initial effort to validate the GPU platform for this application.

In the LEM pedestrians take a near optimal path with as little deviation from an intended or shortest path as possible. A variant of ACO is also used to increase an agent or pedestrian's strategies by which their objective is accomplished. ACO is a population based metaheuristic algorithm. In ACO, foraging ants communicate indirectly through a process called *stigmergy* where they deposit *pheromones* to construct or reinforce a path. *Phermones* evaporate after a point of time if not updated by further deposition. As more ants traverse particular trails, some trails become more prominent than others and eventually only the most efficient trails between food and nest prevails. ACO was first used to provide solutions to the travelling salesman problem (TSP) [8] and has since then been adapted to several applications areas, such as, vehicle routing problems [11], circuit design [11] and protein folding [12]. In our application, we adapt the ant model, considering each pedestrian as an agent that deposits a pheromone trail in the sense of signaling to following agents that the chosen path may be desirable. In this work, a comparison is also provided between the results obtained using the LEM and ACO algorithms.

As each pedestrian functions semi-autonomously on a two dimensional grid, the model is well suited for parallel implementation. As the numbers of agents is also large and their function relatively simple, the type of model is also well suited for processing on Graphics Processing Units (GPUs). GPUs were initially used for graphics purposes but with the introduction of Compute Unified Device Architecture (CUDA) by NVIDIA, GPUs are increasingly being used for general purpose computing. GPUs are now used in numerous applications where there is ample scope to exploit fine grained parallelism. To the best of our knowledge, this is the first data driven based implementation of LEM algorithm for modelling bi-directional pedestrian on a GPU. However, a fair amount of research has already been done in the parallel implementation of ACO on GPU in both task based [13] and data based paradigms [14], [15]. In data driven parallelism, identical threads operate on the data-sets and is better suited for GPU implementations. Using an ACO based model for simulating large bi-directional pedestrian on GPU is also innovative and challenging.

One of the difficulties is dealing with two groups of pedestrians with different goals. Optimization techniques of index mapping and logical operators are utilized to avoid warp divergence, instruction optimization of loop unrolling and also scatter-to-gather transformation is used to avoid atomic operations. Care is also taken towards maintaining the GPU occupancy as well as utilizing different memory options to obtain maximum performance.

The remainder of the paper is arranged as follows. Section II includes a brief description of the ACO algorithm along with the NVIDIA CUDA architecture. Section III provides a description of the pedestrian movement model. Section IV contains a detailed description of the GPU implementation of the model. Section V and VI outline performance analysis and the results of the simulation. Finally we conclude with future plans in section VII.

## II. BACKGROUND

### A. Brief Description of Least Effort Model

The basic mechanism of the LEM [18] permits an agent or pedestrian to make movement decisions that would result in least effort paths. Ideally, agents on a grid would choose the cell nearer to their target rather than any other cells. As shown in Figure 1, a pedestrian located in the central Cell #0 checks the availability of the surrounding cells. The cells are then sorted according to the distances from the target in ascending order. The nearest cell has the lowest rank and the farthest cell has the highest. A random number process is used guide an agent's movement decision.

$$C_i = (1 - n_i)\left(D_{min}/D_i\right) \quad (1)$$
$$D_{min} = Min(D_i), n \in \{0,1\}, D_i \neq 0$$

$D_i$ : Distance of the $i^{th}$ neighboring cell from the target.
$D_{min}$ : Minimum of all the distances.
$n_i$ : 0 for the empty cells and 1 for the occupied ones.

$C_i$s are ranked based on their values in ascending order. The next cell selection process uses a random number from a normal distribution with negative numbers converted to zeroes and the numbers more than the highest $C_i$ are rounded off to the highest $C_i$. This process facilitates a probabilistic decision whereby the agent probabilistically chooses the cell nearest the target.

### B. Brief Description of Ant Colony Optimization

The ACO algorithm was first introduced by Marco Dorigo and his colleagues [9].

Ant System (AS), a variant of ACO, was also proposed by Dorigo [10] and was used to solve the TSP. There are two main stages in AS: *tour construction* and *pheromone update*. In solving the TSP, the process starts with *m* ants placed randomly on the nodes of a graph. There are *n* nodes which are known as cities. The task is to complete the tour by visiting all the cities only once. This tour construction is performed in parallel by all the ants. While moving along the edges, ants deposit pheromones. The amount of pheromones deposited depends on the aggregate pheromones deposited. An ant placed in a particular city chooses the next city to visit depending on a probabilistic rule known as *random proportional rule*.

In the random proportional rule, the transition probability for the $k^{th}$ ant to visit the next city *j* from a city *i* is given by (2).

$$P_{ij}^k(t) = \begin{cases} \frac{[\tau_{ij}(t)]^\alpha [\eta_{ij}]^\beta}{\sum_{l \in N_i^k}[\tau_{il}(t)]^\alpha [\eta_{il}]^\beta}, j \in N_i^k \\ 0, j \notin N_i^k \end{cases} \quad (2)$$

$P_{ij}^k$: Probability of $k^{th}$ ant to go to city *j* while currently at city *i* in the $t^{th}$ step.

$\eta_{ij}$ : Heuristic value from city *i* to city *j*. Generally $\eta_{ij}=1/d_{ij}$, where $d_{ij}$ is the distance between city *i* and *j*.

$\tau_{ij}$ : Amount of pheromone deposited in the path of city *i* and *j*.

α,β: Parameters controlling the relative weight between the pheromone trail and the heuristic value respectively.

$N_i^k$ : Represents the unvisited neighborhood cities that ant *k* could visit at that point of time.

In the pheromone update stage the pheromones are lowered on all the edges by a constant factor.

$$\tau_{ij} \leftarrow (1 - \rho)\tau_{ij} \quad (3)$$

Where $0 < \rho \leq 1$ is the pheromone evaporation rate. After the evaporation, the pheromones are deposited on the visited edges is given in (3)

$$\tau_{ij} \leftarrow \tau_{ij} + \sum_{k=1}^m \Delta\tau_{ij}^k \quad (4)$$

Where $\Delta\tau_{ij}^k$ is the amount of pheromone deposited by ant *k* on the edge between city *i* and city *j*, defined as in (4)

$$\Delta\tau_{ij}^k = \begin{cases} \frac{1}{L_k}, & \text{if the } k_{th} \text{ ant belongs to edge (i, j)} \\ 0, & \text{otherwise} \end{cases} \quad (5)$$

$L_k$ is the length of the tour $T^k$ built by the *k*-th ant and is obtained by the summing up the lengths of the edges in the tour. This AS used for the TSP is modified in our work for pedestrian movement decisions.

### C. Overview of CUDA programming model

With the introduction of Compute Unified Device Architecture (CUDA) [16], [17] by NVIDIA, the use of GPUs for general purpose computing expanded. CUDA provides the means to program NVIDIA GPUs using familiar programming languages as C. Each GPU consists of Streaming Multiprocessors (SM). These SMs are the basic building blocks of GPUs which are basically Single Instruction Multiple Threads (SIMT). The SMs in turn consists of arrays of several stream processors (SP). To program the GPU, massive numbers of threads are arranged into grids and blocks.

GPUs also offer a wide range of memory options. The slowest memory that resides off-chip is the *Global Memory*. This is basically graphics double data rate (GDDR) DRAM whose access time is slowest. The threads within a block communicate with each other using faster on-chip *Shared Memory*. *Registers* are private to each thread, providing the fastest form of memory. *Constant Memory* is a read only type memory where data is cached and could be accessed almost at same speed as *Shared Memory*. Another one such special purpose memory is *Texture Memory*.

The threads arranged into blocks are broken down into groups of 32 threads known as a *warp* which are the smallest unit of thread execution.

### III. MODEL DESCRIPTION

In this section, the simulation models are described. Our objective was to simulate a situation where two groups of pedestrians or agents are placed on the opposite sides of the environment and their target is to reach the other side of the environment. The size of the environment is kept fixed in all the simulations performed. The number of pedestrian varies from one simulation to next, and range from 2K to 100K.

The environment is a 2D grid, divided into cells, each of equal size. The size of the environment chosen is 480x480. The pedestrians are placed at the top and bottom of the environment. Initially the agents on both sides of the environment are placed randomly but kept confined to the pre-defined number of rows. The target of any particular group is the end row of the opposite side. In practice, the objective for the agents is to get as far as possible to the end row of the opposite side until no further movement is possible.

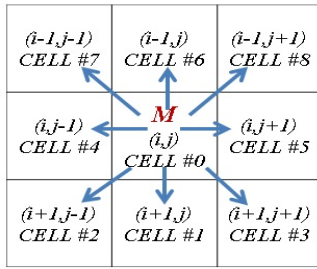

Figure 1. Neighborhood of pedestrian *M* placed in the central cell

At any position within the environment other than the border and corner positions, a pedestrian would always be surrounded by eight (8) surrounding cells. In Figure 1, a pedestrian *M* is placed in the center cell with the coordinate of *(i,j)*. The pedestrian is free to move to any of the surrounding eight (8) cells based on the cell's availability. With the LEM, pedestrians move forward based on the availability of the surrounding cells and their distance from the target (end row of the opposite end). At first, the forward cell of an agent is checked and if found empty then agent chooses that cell as the next one to occupy without carrying out any further calculation. But if the forward cell is found occupied then the agent's movement depends on the rule mentioned in (1). This allows an agent to move forward or to choose the cell nearest to the target most of the time.

Pedestrian simulation is also carried out based on AS rules with certain modifications. For a particular ant or agent, the goal is to reach the other side traversing as few cells as possible. This would provide a real world scenario where pedestrians try to reach their destination with least effort [18], making as minimal deviation from an optimal path as possible. To achieve this, the heuristic value (basically the distance values between cities) is replaced by the heuristic value of the current surrounding cell coordinates and the target which is the last row of opposite end. Pheromone deposition is not a natural phenomenon in humans. However, the concept is used in here to replicate or represent the visual proposition to follow predecessors in a densely populated environment [24].

In the modified ACO a pedestrian also first checks the availability of the front (forward) cell and if found empty, selects cell as the next one to occupy. If the front cell is unavailable, then the pedestrian would consider surrounding cells that are empty. In that process, choosing a particular cell depends on the transition rule mentioned in (2). This is the tour construction phase for the pedestrians.

After all the pedestrians finish with the tour construction phase, the pheromone update stage is performed. The pheromone gets deposited on a particular cell whose value depends on the tour of that particular pedestrian. Pheromone evaporation is also carried out to avoid any local minima situation. This whole simulation is carried out a fixed number of time steps for each simulation.

### IV. IMPLEMENTATION DETAILS

SMs inside the GPUs are the reason for the performance boost in most applications. But the presence of this large array of processors does not make the implementation trivial. There are several factors that need to be kept in mind for performance optimization. Both LEM and modified ACO are used for the pedestrian simulation and GPU optimization factors are of primary concern. Though this is one of the first endeavors of simulating crowd movement using LEM on a GPU utilizing a data driven parallelism, there are some previous implementations of ACO on GPUs [14],[15]. However, there are additional and unique challenges we encountered in our implementation. The primary challenge was to build a data driven model for two groups of pedestrians having completely opposite goals. In [14], the number of ants is considered as the thread block and the numbers of cities are considered as the worker threads. This method is useful when the numbers of cities are large and the numbers of ants are also of considerable size. In our application, the number of cities or neighborhoods for an agent is only eight (8) or fewer. So the previous data driven method is not well suited for our purpose and a whole new model needed to be developed. Other factors such as the occupancy of the GPU, avoiding warp divergence, scatter-to-gather transformations avoiding atomic operation, loop unrolling are also adopted.

The simulation using both LEM and the modified ACO algorithms can be divided into four main stages, each responsible for one kernel. There is also one supporting kernel to carry out the initialization. The GPU that is used here is based on FERMI architecture with a compute capability 2.0[16]. The next subsections describe the five different stages of the simulation in detail (four main stages plus a supporting kernel function).

*a. Data preparation stage*

This stage is actually carried out on the CPU for the LEM and ACO based simulation. This stage is responsible for preparing the data that would be transferred to the GPU. This stage is carried out only once outside the main simulation loop. There are certain constraints that are considered while implementing this stage. As already mentioned, the GPU used in this work has a compute capability of 2.0. Maintaining 100% occupancy [19], the maximum number of threads that could be launched in a single thread block is 256. Tiling technique is used within the kernel functions of subsequent stages and each tile of size 16x16 is chosen. Each thread in our application is assigned to each cell of the environment. So the dimension of threads assigned for a single tile is 16x16. So the total number of threads for a single tile is 256.

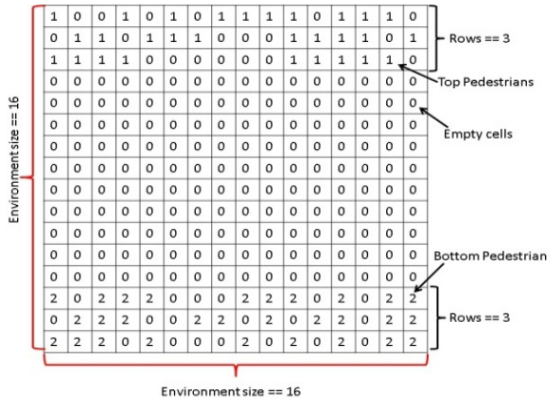

Figure 2a. Sample original environment Matrix *mat* of size 16x16

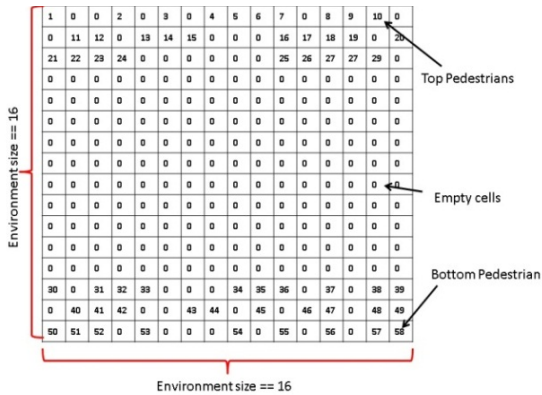

Figure 2b. *index matrix* same as *mat* of size 16x16

| ID | INDEX NO. | ROW | COLUMN | EMPTY | FUTURE ROW | FUTURE COLUMN | FRONT CELL |

Figure 2c. A row of the *property matrix*.

Therefore, an environment size is chosen to be a multiple of size 16. The primary environment matrix *mat* is created, a sample of which is shown in Figure 2a. The agents are placed in the matrix randomly but kept confined to a certain number of rows (e.g. Rows == 3). The top placed agents are labeled 1, the bottom ones are labeled 2 and the empty cells are designated 0. Another matrix similar to the size of *mat* is used to keep the index number of the pedestrians. This matrix is known as the *index matrix* shown in Figure 2b. A *property matrix* is also created in this phase which is responsible for maintaining the properties of the agents. The content of a single row of the property matrix is shown in Figure 2c. Table 1 illustrates the contents in further detail. In this application, the target of the top group of pedestrians is the last row and that for the bottom pedestrians is the first row.

TABLE I. Illustration of the contents of property matrix

| Name | Description |
| --- | --- |
| ID | Identity of the pedestrian, either 1 or 2 |
| INDEX NO | The index value of the matrix from the index matrix. |
| ROW | The present row position of the pedestrian. |
| COLUMN | The present column position of the pedestrian. |
| EMPTY | Unused |
| FUTURE ROW | Holds the future row value that would be decided in the subsequent steps, initially 0 |
| FUTURE COLUMN | Holds the future column value that would be decided in the subsequent steps, initially 0. |
| FRONT CELL | Holds the value of the value that is present in the front cell from the current position. |

The distances of the surrounding cells are pre-calculated in this phase and kept in the *distance matrix*. This *distance matrix* is copied to the constant memory of the GPU, as the values in the matrix remain constant.

One additional matrix is created in the global memory known as the *scan matrix* and initialized to 0. For the LEM it stores the result generated from (1) and for ACO the numerator value of (2) for a particular agent. The total numbers of rows in the *property matrix* and *scan matrix* have 1 more than the total number of pedestrians to avoid the warp divergence within the simulation steps. The purpose of the index number in the *index matrix* is to indicate the row number of *property matrix* and *scan matrix* that is needed to be accessed. The index number starts from 1 and increments up to last pedestrian (58 in Figure 2b). All the other empty cells are 0 in the *index matrix*. The $0^{th}$ row is used to keep the results generated by the calculation of empty cells. Operations on the data stored in *property matrix* and *scan matrix* starts from the $1^{st}$ row, ignoring the $0^{th}$ row. All these matrices are then copied to the global memory of the GPU which are used in the later simulation phases.

For ACO based simulation, another matrix is created to keep the tour length values of the pedestrians, known as the *tour matrix*, and it is initialized to 0. Two separate matrices are used to keep the track of pheromones deposited by the top and bottom pedestrians size same as *mat* matrix.

*b. Initial Calculation Phase*

This phase is launched within the simulation loop as a separate kernel. The *mat* and *index matrix* are copied from global to shared memory using a method known as *tiling* [17] for data re-usage. So these matrices are further subdivided into smaller blocks of size 16x16. So the total number of threads launched is equal to the total number of cells of the *mat* matrix and each block consists of 256 threads, maintaining 100% occupancy.

The purpose of this kernel is to determine the availability of surrounding cells and then obtain their distances to calculate (1) or numerator of (2). For the agents placed on the corner and border cells, a memory violation could occur as the assigned thread tries to access some of the neighboring cells. For example, when the thread assigned for the red cell in Figure 3, tries to access neighboring Cell #2,4,6,7 and 8 shown in Figure 1 there would be memory violation as those elements does not belong to the same tile or are not present at all. So the shared memory matrix is declared of size 18x18. The elements from the same block are called *internal elements* and the cells marked in blue are the elements from the immediate next block are called *halo elements* as shown in Figure 3. The loading of the halo elements usually involves a number of thread divergences. There are 32 threads involved for first 2 rows that are marked in green (and one cell in red). Through index mapping techniques, this whole warp is used to load the *halo elements* marked in blue and avoids the thread divergence while loading the elements.

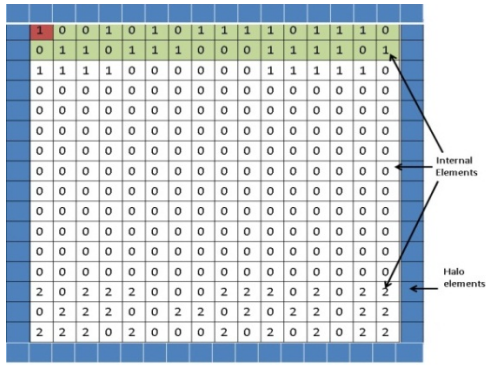

Figure 3. Loading of *Halo Elements* from global to shared memory

In this stage, calculation of (1) is carried out for the LEM based simulation. This operation is carried out for the occupied central cell. The threads consider only empty neighborhood cells. This calculation is accomplished using the index operation and logical operators avoiding any warp divergence. As shown in Figure 1, agents are placed in central cell denoted Cell #0. If the agent was placed initially on the top portion of the environment then Cell #1 will always be the cell with a minimum distance to the other side if unoccupied. Cell #2 and Cell #3 have equal distances from the target which is greater than Cell #1 but less than remaining. The same process is carried out for the remainder of the unoccupied adjacent cells. For the bottom placed agents this same calculation would be the opposite. All these distances are fetched from the *distance matrix* for the neighboring empty cells and are stored in the *scan matrix*. In the LEM based simulation, each row of the *scan matrix* stores the distances in the increasing order of value. Index number from the *index matrix* is used to access the row of the *scan matrix*.

For ACO based simulation, the calculation of the numerator of (2) is carried out in the same manner. But to achieve this, two separate global pheromone matrices are loaded into a single local pheromone matrix. The local pheromone matrix has 36 rows and 18 columns. First 18 rows are used for top agents and next 18 rows for bottom agents. A pedestrian label (i.e. 1 and 2) is used to access proper cells, avoiding warp divergences. Now the numerator of (2) is calculated and gets stored in *scan matrix* in the same fashion as carried out for LEM based simulation. The row to access scan matrix is obtained from the index matrix. The empty cells are marked as 0 in the index matrix. So for both types of simulation, the $0^{th}$ row of the *scan matrix* is used to store the values generated from the threads assigned to the empty cells and would be ignored in the subsequent operation. Care was also taken towards calculating the maximum shared memory and the number of registers that could be used without endangering the 100% occupancy.

*c. Tour Construction Phase*

In this phase, agents decide to move to neighboring empty cells. The operations in this phase are almost identical for both the LEM and ACO based simulation with subtle differences. For both types of simulation, the *scan matrix* is first loaded in the local matrix. The total number of threads launched is equal to 8 times the total number of agents in the environment. The threads are grouped into blocks each containing 32 rows and so each block has 256 threads maintaining 100% GPU occupancy. Values from the $1^{st}$ row are loaded to the local matrix in shared memory as the $0^{th}$ row contains the data generated from the empty cells.

Particularly for the LEM based simulation, the local *scan matrix* contains the distances of the surrounding empty cells for a particular agent in ascending order. So the first element of each row of the matrix has the lowest value (Cell #1 for top placed agent and Cell #6 for bottom placed) and the last element has the highest value (Cell #8/Cell #7 for top placed agent). The first 4 rows of each block contain 32 threads forming a warp. These 4 rows are involved in a reduction operation adding the values of each row avoiding warp divergences. Using the CURAND library [20], a random number is generated from the normal distribution and then rounded off as outlined in [18]. All the approximations of values are carried out using logical operators again avoiding warp divergence. We modified the LEM of [18] as in our case forward movement is given the highest priority. So if the forward cell is empty agents decide to move to the forward cell in the next step.

For ACO based simulation, the primary objective of this phase is to calculate (2) and choose a destination cell for the next step by all the agents. This is performed using data brought in the local matrix from the global *scan matrix*. A reduction operation is also performed in this simulation to obtain the denominator using first 4 rows of each block forming a warp. A random number is generated using the CURAND library but without any rounding. The random number determines the cell to be selected, with the front (forward) cell having the highest preference. If the front cell is empty, then the pedestrian decides to move forward immediately. Otherwise the agent decides to move to a cell chosen probabilistically.

For both types of simulation, after the decision is made, the selected future cell coordinate is stored in the FUTURE ROW and FUTURE COLUMN in the *property matrix*.

*d. Agent Movement Stage*

This phase is responsible for the agent movement and the matrices are updated. For both types of simulation, the configuration of threads and blocks launched in this stage is similar to the configuration in the *Initial Calculation Phase*. Each block has a configuration of 16x16 threads. Similar to method illustrated in Figure 3, *mat* and *index matrix* are copied to the local shared memory matrix variable of size 18x18. Specifically the ACO based simulation, both top and bottom *pheromone matrix* are copied from the global memory to local shared memory. Both pheromone matrices are copied to a single local pheromone matrix with 32 rows (top 16 rows for top agents and the remaining 16 rows for bottom agents) and 16 columns. The agent label is used to access the proper portion. Each thread is responsible for each *internal element* of the loaded block.

In the ACO simulation, the pheromone is reduced similar to (3) at first. This operation is trivial as all the threads operating on the inner elements reduce the pheromones in the local memory at the same rate.

For both kinds of simulations, in this phase the threads assigned to the empty cells participate in the calculation in contrary to the *initial calculation phase*, where the threads for occupied cells participated in calculation. In Figure 4, a typical situation is demonstrated. The cell identifiers are marked in red and the surrounding pedestrians are marked in black. Arrows illustrates the movement of the surrounding pedestrian to the central cell with coordinate *(i,j)*. This sort of race condition could be avoided by using the CUDA *atomic operations* [16], [17]. But an atomic operation serializes an application and thus increases computation time. As such a more advanced technique known as *scatter-to-gather* [21] is used to make the movements decisions.

The thread assigned to the cell with coordinate *(i,j)*, checks the number of agents in neighborhood cells and decides to select a cell randomly in the next step. This information is available in the FUTURE ROW and FUTURE COLUMN of the surrounding agents. In Figure 4, the thread for *(i,j)* after counting gives an output of 5 as agents in cell 0,1,3,5 and 6 moves to central cell in next step. But Figure 4 shows only one type of situation among all the other possible combinations. There are threads operating on the other cells of internal elements of the local *mat* matrix that may not be empty (as in cells 0,1,3,5 and 6). One more situation that could arise is when no surrounding agents decide to move to the central cell even though it is empty. In all these cases the result after counting the neighboring cells should be 0. Considering all the conditions, the counting is performed using logical operators, thus avoiding any warp divergence. Only the cells generating a count greater than 0 are considered and the rest are neglected. Based on the count result, a random number is generated from the CURAND library uniform distribution. The number indicates the cell number whose content is to be exchanged with the central cell. The ROW and COLUMN in the *property matrix* is also updated with the coordinate of the central cell. The row of the *property matrix* to be accessed is indicated by the index number from the *index matrix*. Ultimately the value present in the *index matrix* is also exchanged in the same way as performed in *mat* matrix.

For ACO driven simulation, the pheromone value of the central cell also need to be increased. Cells from the proper *pheromone matrix* need to be increased. However, before incrementing pheromones, the tour length is added in the *tour length* matrix. The index number of the chosen cell is used to indicate the row to access in the *tour length* matrix. The value that is to be added to the present value of the tour length of that agent is stored in constant memory, reducing the calculation time. Through this process, all the updating is performed without involving any atomic operations for both types of simulation.

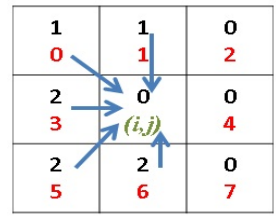

Figure 4. An illustration of race condition occurrence

*e. Supporting Kernel Funtions*

There is one additional kernel function launched to initialize the values of the *scan matrix* and the FUTURE ROW and FUTURE COLUMN of the *property matrix*. After the initialization, control is returned to *Initial Data Calculation Phase* and simulation is carried out up to a pre-defined number of time steps.

V. PERFORMANCE ANALYSIS

In this section, we compare and analyze the performance between the GPU and CPU. We also perform a comparison between the execution time of ACO and LEM based simulation on GPU. The NVIDIA GeForce GTX 560ti GPU is used with Fermi architecture. On the CPU side, Intel Core i7-930 processor is used. The specifications of both hardware units are provided in TABLE I.

TABLE I. Hardware specifications of CPU and GPU used

| Attributes | CPU and GPU hardware specifications | |
|---|---|---|
| | CPU | GPU |
| Manufacturer | Intel | Nvidia |
| Model | Core i7-930 | GeForce GTX 560ti |
| Processor Cores | 4 | 448 |
| Clock Frequency (GHz) | 2.8 | 1.464 |
| L1 Cache size | 32 KB + 32 KB | 16 KB + 48 KB (shared memory configurable) |
| L2 Cache size | 256 KB/ core | 768 KB |
| L3 Cache size | 8 MB | Not available |
| DRAM Memory | 6 GB DDR3 | 1.25 GB GDDR5 |

The programming is performed on Microsoft Windows 7, and Microsoft Visual Studio 2010 is used to perform the programming on the CPU side. The NVIDIA CUDA compilation tools release 5.0 is used on the GPU side. The time spent solely for simulation on GPU is measured using the *cudaevent* functions. On the CPU side, the *time* function is used. Data are recorded into text files and MATLAB is used for plotting.

In Figure 5a-c, the X-axis denotes the number of total agents in the environment. It starts with 2560 agents and ends with 102,400 agents. For all simulation the numbers of steps are kept constant at 25,000. In Figure 5a, the Y-axis denotes the execution time of the ACO and LEM based model in seconds.

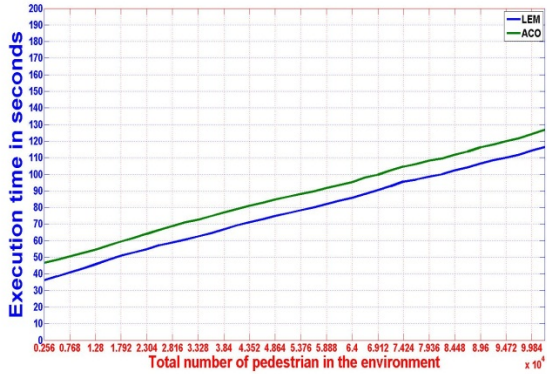
Figure 5a. Execution time for ACO and LEM in seconds

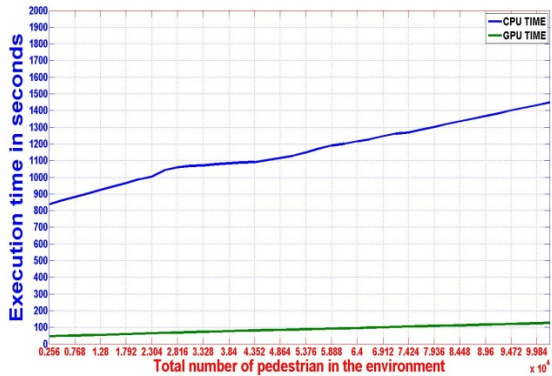
Figure 5b. . Execution time for CPU and GPU in seconds

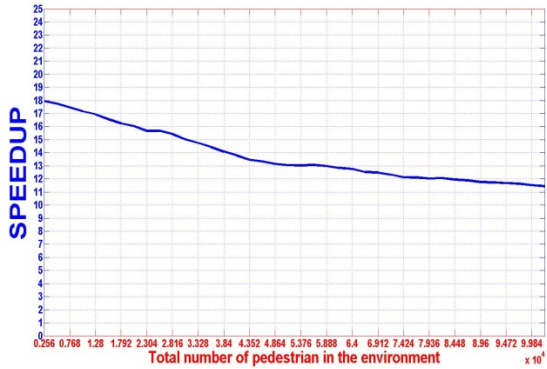
Figure 5c. Speedup graph comparing CPU and GPU based ACO simulation

The execution time of the ACO and LEM are found to be almost same. There is a marginal increase of 11% in the execution time of ACO compared with the LEM based model because of the additional operations. Figure 5b depicts the time measurement for ACO based simulation on CPU and GPU in seconds and Figure 5c shows the speedup factor. In Figures 5b, for 2560 agents the time requirement for the completion of simulation is 46.66 seconds on GPU whereas it is 837.5 seconds on the CPU. We gained a significant speedup factor of 18x. After that, a steady decrease in the speedup is observed as the number of pedestrian increases. In the final simulation for 102,400 pedestrians, a speedup factor slightly higher than 11x is obtained. The time required to complete the simulation is 1449 seconds on the CPU and 126.7 seconds on the GPU.

## VI. RESULTS

The quality of the simulation results obtained from the GPU implementation is assessed by comparing it with the CPU counterpart. In our case, benchmarking is not possible as was done in [14] that used TSPLIB [22]. Comparing the solution obtained from CPU and GPU is a viable way to begin to establish consistency of the implementation, as part of an overall validation of the approach. We also compared the result obtained from the LEM based simulation and its ACO counterpart, both implemented on the GPU

We define *throughput of pedestrians* as the number of pedestrians able to cross the environment and reach the other side and the number of time steps required. As mentioned previously, initial pedestrian positions, though random, are kept restricted up to certain number of rows (e.g. 3 in Figure 2a). The throughput is recorded as the number of pedestrians able to cross over to the opposite side (e.g. $14^{th}$ row in the opposite end for top positioned pedestrians). The size of the environment is chosen to be of size 480x480 for all the models and the simulations carried out. The results obtained from all the simulations are shown in Figure 6a and in Figure 6b. The total number of pedestrians in the environment starts with 2560 (1280 in each side), and is increased by 2560 pedestrians for each simulation instance up to 102,400 pedestrian in total (51,200 on each side). However, beyond the total population of 51,200, the throughput of pedestrians becomes insignificant. Here the result is shown for first 20 simulations when the pedestrian total reaches to 51,200. All the simulations are repeated 10 times and then the results are averaged. The number of simulation steps is 25,000 which has been kept fixed in all the scenarios.

Figure 6a, depicts the throughput of the agents in ACO and LEM based simulation, both implemented on GPU. The Y-axis is the total number of agents that are able to cross to the other side of the environment. The X-axis is the simulation number (agent density) after averaging over 10 simulations run. The simulation starts with a total of 2560 agents and increases at 2560 agents in the next simulation run. As mentioned, throughput up to 20th population density simulation is considered where total population is 51,200. After this, the pedestrians are either unable to cross the other side within the fixed amount of time steps or the number that are able to cross is insignificant (total gridlock). From Figure 6a it is evident that for first 9 simulation scenarios, the throughput for both the ACO and LEM based simulation

is effectively the same. However, in the 10th simulation run, we can see a decrease in throughput for the LEM based model compared with the ACO based model, both simulated on a GPU. There are total of 25,600 agents in the environment. In the LEM based model, 17,417 numbers of agents are able to cross in 25,000 time steps whereas for ACO based model, 25,600 agents are able to cross in the same number of time steps. Maximum throughput for the ACO driven simulation is observed in the 11th simulation scenario, with a total number of 28,160 pedestrians in the environment. For the ACO and LEM, the agent throughputs were 28,160 and 5,272 respectively. In all the subsequent simulation scenarios, the throughput of ACO based model is higher than the LEM based model. Considering all the 20 simulation scenarios, there is an overall increase of 39.6% in the throughput using the ACO based model over that LEM based model. The higher throughput obtained using the ACO-driven simulation is interpreted as a better result than the LEM-based simulation as a consequence of the pedestrians having greater agency or decision making capacity.

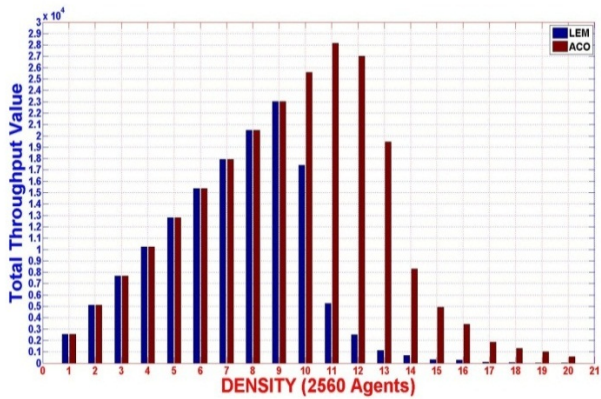

Figure 6a. Graph comparing the throughput of pedestrian for LEM and ACO model implemented on the GPU

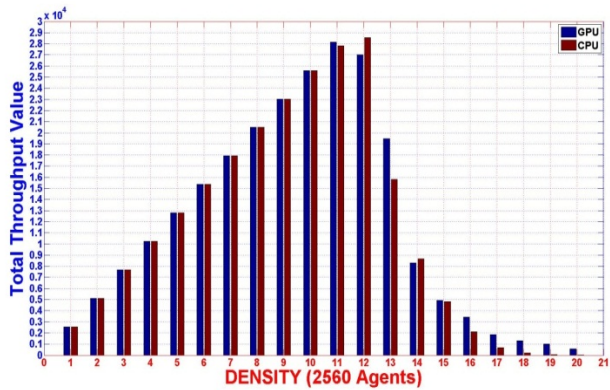

Figure 6b. Comparison of throughput on CPU and GPU for ACO based pedestrian simulation

Figure 6b depicts the comparison between the throughputs of ACO based simulation on the GPU and CPU. The results are obtained after carrying out each simulation scenario 10 times and then averaging the results. The throughputs for both the CPU and GPU based simulation are almost identical. Subtle differences are observed from the 11th simulation run onwards. However, apart from the 12th simulation run (with total of 30,720 agents) the throughput for the GPU based simulation is comparable to the CPU based simulation. To compare GPU performance with CPU performance we can see that we can model this scenario by a binomial generalized linear model (glm) [25], where the probability that as agent crosses over to the other side is modeled with respect to the different number of agents and an indicator for the simulation run being run on either the CPU or GPU. We have 40 different simulation scenarios each with variable agent sizes (1st scenario starting with 2560 agents and 102,400 agents in the 40th scenario). We suppress the first 10 scenarios and the last 10 scenarios for comparing CPU and GPU. For most of the first few scenarios, all agents cross over in both cases, while in most cases in the last 10 scenarios none of them actually cross over. Test for the statistical significance of the difference of performance between CPU and GPU in this model used a t-test, and did not provide sufficient evidence to reject the test (p-value=0.6145). Hence the throughput performance generated by the GPU is similar to the one by CPU.

VII. CONCLUSION AND FUTURE WORK

In this paper we have demonstrated that using ACO based simulation produces higher throughput, considered to be a better result when compared to the LEM based model, both run on a GPU. In the LEM model, agents tries to move only taking the least-distance path towards the target, most of the time making as little deviation as possible forms a near optimal path. In the modified ACO driven simulation, agents have some additional intelligence. In this type of simulation agents move by following their predecessor using a pheromone trail in conjunction with a least effort mode. Results produced by our modified ACO based simulation model surpassed the LEM based model by 39.6% overall. In effect, as evident from Figure 6a, the LEM and ACO are virtually identical when the pedestrian density is low, the ACO provides for more optimal paths when the density is medium, and when highly congested neither the LEM nor ACO offer a means for pedestrian movement. This significant increment in throughput is obtained by a minimal increment of 11% in the execution time in the ACO based simulation compared to its LEM counterpart.

The simulation results also indicate that the results of an ACO model simulated on both the CPU and GPU are similar to one another, adding credibility to the quality of the solution obtained from the GPU platform. This work demonstrated the techniques adopted to avoid warp divergences and achieve substantial speedup with a large number of pedestrians in the simulations. Sophisticated index mapping techniques were adopted, such as loading of the edge elements from the neighbor tiles which could be helpful in other applications such as image processing.

Utilizing data parallelism and using a metaheuristic algorithm to emulate social dynamics is still in its preliminary stage and needs further investigation. While implementing the above algorithm, many simplifying assumptions were made. The velocity and size of the pedestrians are kept constant in all the simulations performed. In the next phase, one of our challenges will be to introduce more complexity into pedestrian characteristics while maintaining data parallelism. Another objective is to introduce a panic alarm to emulate some sort of crisis situation. Furthermore, separating the scanning ranges and moving ranges of the pedestrians would be an interesting feature to introduce and simulate. Currently, the scanning range and the movement range of the pedestrians are confined only to the neighboring cells. Increasing the scanning range as well as the movement range and using different values for scanning and moving ranges to make decisions would be more practical and would add realism to the simulation. On the hardware side, several enhancements can be considered. The GPU that is used has Fermi architecture. Using Kepler Architecture [23] with advanced features would add to the performance. CUDA *streams* on Fermi GPU actually operate in a queue. However, on Kepler GPUs, they can be launched as actual separate *streams* in parallel. Using advanced GPUs will also provide more CUDA cores along with more memory for all stages.

Notwithstanding future enhancements, the present work with its corresponding techniques can be considered a novel approach to using nature inspired metaheuristics for emulating a large number of pedestrians on GPUs utilizing data parallelism.